\documentstyle[epsfig,preprint,aps]{revtex}           
\begin{document}       
\tighten
\def\bea{\begin{eqnarray}}          
\def\eea{\end{eqnarray}}          
\def\beas{\begin{eqnarray*}}          
\def\eeas{\end{eqnarray*}}          
\def\nn{\nonumber}          
\def\ni{\noindent}          
\def\G{\Gamma} 
\def\D{\Delta}         
\def\d{\delta}          
\def\l{\lambda}          
\def\g{\gamma}          
\def\m{\mu}          
\def\n{\nu}          
\def\s{\sigma}          
\def\tt{\theta}          
\def\b{\beta}          
\def\a{\alpha}          
\def\f{\phi}          
\def\fh{\hat{\phi}}          
\def\y{\psi}          
\def\z{\zeta}          
\def\p{\pi}          
\def\e{\epsilon}           
\def\ve{\varepsilon}
\def\cd{{\cal D}}          
\def\cl{{\cal L}}          
\def\cv{{\cal V}}          
\def\cz{{\cal Z}}          
\def\pl{\partial}          
\def\ov{\over}          
\def\~{\tilde}          
\def\rar{\rightarrow}          
\def\lar{\leftarrow}          
\def\lrar{\leftrightarrow}          
\def\rra{\longrightarrow}          
\def\lla{\longleftarrow}          
\def\8{\infty} 
\def\pls{\partial\!\!\!/}
\def\bs{b\!\!\!/}
\def\ps{p\!\!\!/}
\def\qs{q\!\!\!/}
\def\ls{l\!\!/}
\def\rs{r\!\!\!/}
\def\ks{k\!\!\!/}
\def\As{A\!\!\!/}
\def\yb{\bar{\y}}
\def\Ds{D\!\!\!\!/}    
\newcommand{\fr}{\frac}       

\title{Lorentz and CPT Violating Chern-Simons Term in the Formulation of 
Functional Integral}
          
\author{J.-M. Chung\footnote{Electronic address: chung@ctpa03.mit.edu}}
\address{Center for Theoretical Physics,
Massachusetts Institute of Technology,\\
Cambridge, Massachusetts 02139}
       
\date{MIT-CTP-2853,~~~~ April 1999}                  
\maketitle              
\draft              
\begin{abstract}           
\indent           
We show that in the functional integral formalism 
the (finite) coefficient of  the induced, Lorentz- and CPT-violating 
Chern-Simons term, arising from the Lorentz- and CPT-violating fermion sector, 
is undetermined.
\end{abstract}              
                   
\pacs{PACS number(s): 12.20.-m, 11.30.Cp}           
         
In recent work \cite{ck}, the following question has been posed: 
is a Lorentz- and CPT-violating Chern-Simons term induced by 
the Lorentz- and CPT-violating term $\bar{\y}\bs\g_5\y$ ($b_\m$ a constant 4-vector)
in the conventional Lagrangian of QED. 
In the paper by Jackiw and Kosteleck\'{y} \cite{jk}, this problem has been  
discussed. Their results depend on whether one uses a nonperturbative formalism or a 
perturbative formalism. In a nonperturbative formalism, radiative corrections arising
from the axial vector term in the fermion sector induce a definite and nonzero 
Chern-Simons term, while when a perturbative formalism is used, radiative corrections 
are finite but undetermined. 

The purpose of this note is to view this 
undeterminicity\footnote{See also Refs.~\cite{jw,ch}} of the finite radiative 
corrections in the functional integral formalism.

Let us consider the following functional integral
\bea
Z(A)=\int d\bar{\y} d\y \exp[iI(A)]\;,\label{za}
\eea
where 
\beas
I(A)=\int d^4x[i\bar{\y}\Ds\y-m\bar{\y}\y-\bar{\y}\bs\g_5\y]\;.
\eeas
The gauge field $A$ in the covariant derivative $\Ds$ is external.
By changing the field variables
\beas
\y(x)&\rar& \exp[i\a(x)\g_5]\y(x)\;,\nn\\
\bar{\y}(x)&\rar& \bar{\y}(x) \exp[i\a(x)\g_5]\;,
\eeas
we see that the integration measure of (\ref{za}) changes by 
\beas
 d\bar{\y} d\y\rar d\bar{\y} d\y\exp\biggl[-i\int{\a(x)\ov 8\p^2}
\,^*\!F^{\m\n}(x)F_{\m\n}(x)\biggr]\;,
\eeas
where this form for the anomaly is  obtained when $\yb\Ds\y$ is defined 
in a gauge-invariant manner. The action changes into
\bea
I(A)\rar\int d^4x\Bigl[i\bar{\y}\Ds\y-\pl_\m\a J^\m_5 
-m\bar{\y}e^{2i\a(x)\g_5}\y-\yb \bs \g_5 \y \Bigr]\;,\label{ia}
\eea
where the second term ($\pl_\m\a J^\m_5$) in the integrand of (\ref{ia}) 
comes from transformation of the gauge invariant kinetic 
term, hence the axial vector current $J^\m_5$ {\em is} gauge invariant, 
and does not necessarily equal to
$\yb \g^\m \g_5 \y$, which need not be gauge invariant. 
However, we must insist that $\int d^4x \bar{\y}\g_\m\g_5\y$ is  gauge invariant,
i.e., the density need not be gauge invariant but its space-time integral
--- the action --- is gauge invariant. Since the Chern-Simons term 
${1\ov 2}\e^{\m\a\b\g}F_{\a\b}A_\g=\,^*\!F^{\m\n}A_\n$ behaves precisely 
in this same way under gauge transformation, we  may use it to represent 
gauge non-invariant portion of $\bar{\y}\g^\m\g_5\y$. Thus we write
\beas
\yb\g^\m\g_5\y=J^\m_5-c\,^*\!F^{\m\n}A_\n\;,
\eeas
where $c$ is an unknown constant.
By choosing $\a(x)=-x^\m b_\m$, $J_5^\m$ cancels and we arrive at
\beas
Z(A)=e^{-i\int{d^4x\ov 4\p^2}b_\m\,^*\!F^{\m\n}A_\n}
e^{ic\int d^4x b_\m\,^*\!F^{\m\n}A_\n}
\int d\bar{\y} d\y \exp\biggl[i\int d^4x\Bigl[i\bar{\y}\Ds\y
-m\bar{\y}e^{2i\g_5x^\m b_\m}\y\Bigr]\Biggr]\;.
\eeas

Now let us calculate the vacuum polarization for the theory governed by the 
action in the functional integral. The propagator is given as
\beas
G(x,y)={i\ov i\pls-me^{2i\g_5 x\cdot b}}\d (x-y),
\eeas
which to first order in $b$ becomes
\beas
&&\!\!\!\!\!{i\ov i\pls-m-2im\g_5x^\a b_\a}\d (x-y)\nn\\
&\approx&S(x-y)-i\int S(x-z)2im \g_5 z^\a b_\a S(z-y) d^4 z\nn\\
&\equiv&S(x-y)+\D G(x,y),
\eeas
where $S(x-y)$ is the free propagator ${i\ov i\pls-m}\d (x-y)$. 
This decomposition of the propagator splits the vacuum poraization tensor 
into three parts:
\beas
\Pi^{\m\n}=\Pi^{\m\n}_0+\Pi^{\m\n}_b+\Pi^{\m\n}_{bb}\;.
\eeas
The first term $\Pi^{\m\n}_0$ is the usual lowest-order vacuum polarization tensor
of QED, and the last term $\Pi^{\m\n}_{bb}$ is at least quadratic in $b$.
Our main concern here is to calculate the middle term $\Pi^{\m\n}_b$, which is 
linear in $b$. It is expressed as
\beas
\Pi^{\m\n}_b(x,y)&=&{\rm tr~} [\g^\m S(x-y)\g^\n \D G(y,x)]
+{\rm tr~}[\g^\m \D G(x,y) \g^\n S(y-x)]\nn\\
&\equiv&\Pi^{\m\n}_{b1}(x,y)+\Pi^{\m\n}_{b2}(x,y)\;.
\eeas
The Fourier transform of $\Pi^{\m\n}_{b1}(x,y)$ is readily obtained as:
\beas
\Pi^{\m\n}_{b1}(p,q)&=&\int d^4 x\, d^4 y\, e^{-ipx} e^{iqy}\,\Pi^{\m\n}_{b1}(x,y)\nn\\
&=&\int d^4 x\, d^4 y\,d^4z\, e^{-ipx} e^{iqy}{\rm ~tr~} [\g^\m S(x-y)\g^\n 
(-i)S(y-z)2im \g_5 z^\a b_\a S(z-x)]\nn\\
&=&-2imb_\a{\rm ~tr\,} \biggl[\int d^4 x\, d^4 y\,d^4z{d^4l\ov (2\p)^4}
{d^4r\ov (2\p)^4}{d^4k\ov (2\p)^4}e^{i(-p-l+k)x}e^{i(l-r+q)y}e^{i(r-k)z}\nn\\
&&~~~~~~~~~~~~~~~
\times\g^\m {1\ov \ls-m}\g^\n {1\ov \rs-m}\g_5 z^\a {1\ov \ks-m}\biggr]\;.
\eeas
After carrying out $x$, $y$, $z$, $r$, and $k$ integrations, 
we are left with a simple momentum integration:
\beas 
\Pi^{\m\n}_{b1}(p,q)&=&
-2imb_\a{\rm ~tr\,} \biggl[\int {d^4l\ov (2\p)^4}
\pl^\a_q\d(q-p)\g^\m {1\ov \ls-m}\g^\n {1\ov \ls+\qs-m}
\g_5{1\ov \ls+\ps-m}\biggr]\nn\\
&=&2imb_\a\d(q-p){\rm ~tr\,} \biggl[\int {d^4l\ov (2\p)^4}
\g^\m {1\ov \ls-m}\g^\n\pl^\a_q {1\ov \ls+\qs-m}
\g_5{1\ov \ls+\ps-m}\biggr]\;.
\eeas
Note that in spite of the explicit $x$-dependence in the action, the vacuum 
polarization is translation invariant (proportional to $\d (p-q)$ in momentum
space).

Using the identity
\beas
\pl^\a_q {1\ov \ls+\qs-m}=-{1\ov \ls+\qs-m}\g^\a{1\ov \ls+\qs-m}\;,
\eeas
we have
\beas
\Pi^{\m\n}_{b1}(p,q)&=&
-2imb_\a\d(q-p){\rm ~tr\,} \biggl[\int {d^4l\ov (2\p)^4}
\g^\m {1\ov \ls-m}\g^\n{1\ov \ls+\qs-m}\g^\a{1\ov \ls+\qs-m}
\g_5{1\ov \ls+\ps-m}\biggr]\;,
\eeas
which can be written as
\beas
\Pi^{\m\n}_{b1}(p,q)&=&
2imb_\a\d(q-p){\rm ~tr\,} \biggl[\int {d^4l\ov (2\p)^4}
\g^\m {1\ov \ls-m}\g^\n{1\ov \ls+\ps-m}\g^\a\g_5{1\ov (l+p)^2-m^2}
\biggr]\nn\\
&\equiv&\d(p-q)b_\a\~{\Pi}^{\m\n\a}_{b1}(p)\;.
\eeas
If we observe that $\Pi^{\m\n}_{b2}(x,y)=\Pi^{\n\m}_{b1}(y,x)$, then we readily
obtain that the momentum conserving Fourier transform of $\Pi^{\m\n}_{b2}(x,y)$ is 
given by
\beas
\~{\Pi}^{\m\n\a}_{b2}(p)=\~{\Pi}^{\n\m\a}_{b1}(-p)\;.
\eeas
The calculation of the integral determining 
$\~{\Pi}^{\m\n\a}_{b1}(p)$ can be found
in Ref.~\cite{jk}:
\beas
\~{\Pi}^{\m\n\a}_{b1}(p)={-i\e^{\m\n\a\b}p_\b\ov 4\p^2}{\tt\ov \sin\tt}\;,
\eeas
where
$\tt\equiv 2 \sin^{-1}(\sqrt{p^2}/2m)$ and $p^2<4m^2$.
Therefore we obtain the induced, parity violating, non-local action, 
which agrees with \cite{jk},
\beas
{1\ov 4\p^2}\int{d^4p\ov (2\p)^4}b_\m\,^*\!F^{\m\n}(p)
\biggl[{\tt\ov \sin \tt}-1+4\p^2 c\biggr]A_\n(-p)\;.
\eeas
Since $(\tt/\sin \tt)_{p^2=0}=1$, we get the Chern-Simons term with an undetermined
strength $c$. This completes our argument.

We have shown in the functional integration formalism that 
the (finite) coefficient of  the induced, Lorentz- and CPT-violating 
Chern-Simons term 
arising from the Lorentz- and CPT-violating fermion sector is undetermined.
Our result is basically a consequence of the weaker condition for gauge 
invariance:
since $\bar{\y}\g_\m\g_5\y$ does not couple to any other field, 
physical gauge invariance
is maintained provided $\bar{\y}\g_\m\g_5\y$ is  gauge invariant at zero 
4-momentum \cite{jk}.

\acknowledgements
I would like to thank Professor R. Jackiw for enlightening
discussions and to acknowledge the Center for Theoretical 
Physics, MIT, for the warm hospitality. This work was supported in part by 
the Korea Science and Engineering Foundation, 
and in part by the United States Department of Energy under grant number 
DF-FC02-94ER40818.

\end{document}